\documentclass[10pt,twocolumn]{IEEEtran_v15}
\makeatletter
\def\ifundefined{\@ifundefined}
\makeatother

\usepackage{epsfig}

\begin{document}

\setlength{\evensidemargin}{-0.35in}
\setlength{\oddsidemargin}{-0.35in}
\setlength{\textheight}{9in}
\setlength{\topmargin}{0in}
\setlength{\headheight}{0in}
\setlength{\headsep}{0in}
\centerfigcaptionstrue
\title{An RF Circuit Model for Carbon Nanotubes}

\author{P.J. Burke\\
Integrated Nanosytems Research Facility\\
Department of Electrical and Computer Engineering\\
University of California,
Irvine\\
Irvine, CA, 92697-2625}
%\author{P.J. Burke}
%\address{INRF}

\maketitle

% we have to get rid of the headings on the title page
\thispagestyle{empty}
% and on all pages that follow
\pagestyle{empty}

\begin{abstract}
We develop an rf circuit model for single walled carbon nanotubes
for both dc and capacitively contacted geometries.
By modeling the nanotube as a
nano-transmission line with distributed kinetic and magnetic
inductance as well as distributed quantum and electrostatic
capacitance, we calculate the complex, frequency dependent impedance
for a variety of measurement geometries. Exciting voltage waves on the
nano-transmission line is equivalent to directly exciting the yet-to-be
observed one dimensional plasmons, the low energy excitation of a
Luttinger liquid.\footnote{This work has been submitted to the IEEE for possible
publication.}
\end{abstract}

\section{Introduction}

Our goal in this paper is
to describe an rf circuit model for the effective electrical 
(dc to GHz to THz) properties of carbon nanotubes. 
While we restrict our
attention to metallic single walled nanotubes, 
the general approach can be used to describe
semiconducting carbon nanotubes, multi-walled carbon nanotubes, quantum
wires in GaAs heterostructures\cite{Auslaender}, and any other system
of one-dimensional interacting electrons\cite{Fisher}.
An additional goal of this paper is to describe a technique that can be
used to directly excite 1d plasmons in carbon nanotubes using 
a microwave signal
generator. This technique was recently applied to measure collective
oscillations (plasmons) in a two-dimensional electron gas, including
measurements of the 2d plasmon velocity, as well as the temperature
and disorder dependent damping\cite{BurkeAPL}. 
The high frequency circuit model developed herein
may have direct applications in
determining the switching speed of a variety of nanotube based
electronic devices.

In our recent 2d plasmon work, we suggested a transmission-line
effective circuit model to relate our electrical impedance measurements to the
properties of the 2d plasmon collective 
excitation\cite{BurkeAPL,BurkeUnpublished,PeraltaAPL,PeraltaThesis}.
There, we measured the kinetic inductance of a two-dimensional
electron gas, as well as its distributed distributed 
electrostatic capacitance to
a metallic ``gate'' by directly exciting it with a microwave voltage.
The distributed capacitance and inductance form a transmission line,
which is an electrical engineer's view of a 2d plasmon.

Since then, the transmission-line description has
been discussed in the context of both
single-walled\cite{BockrathThesis} and
multi-walled\cite{Tarkiainen,Sonin} carbon nanotubes. In
reference~\cite{BockrathThesis},
 by considering the Lagrangian of a one-dimensional
electron gas (1DEG), an expression for the quantum capacitance (which was not
important in our 2d experiments) as well as the kinetic inductance of
a SWNT is derived. In reference~\cite{Tarkiainen,Sonin} the
tunnel conductance at high voltages is related to electrical
parameters (the characteristic impedance) of the
transmission line in a multi-walled nanotube. In both of these
discussions, the distributed inductance and capacitance per unit length
form a transmission line, which is an electrical engineer's
description of a 1d plasmon. 

In what follows we present an rf circuit model based on the
transmission line properties of a carbon nanotube.
We use this model to calculate the nanotube
dynamical impedance (real and imaginary) as a function of frequency,
as well as the ac damping and wave velocity. 
We discuss possible practical consequences\cite{Dyakonov2001} 
of the results in nanotube electronic and micro/nano-mechanical 
high-frequency circuits. A more detailed discussion of the rf circuit
model can be found in our recent manuscript\cite{BurkeNanotube}.
A description from a theoretical physics point of view
(complementary to the circuit description discussed here)
can be found in reference~\cite{Blanter}, and references therein.

\begin{figure}
\epsfig{file=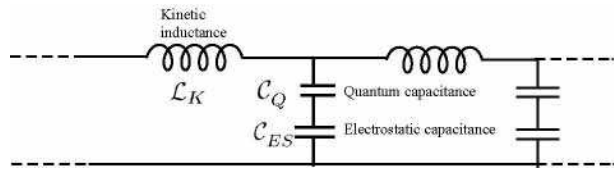}
\caption{Circuit diagram for 1d system of spinless electrons. Symbols are defined per unit length.}
\label{fig:spinlesscircuit}
\end{figure}

\begin{figure}
\epsfig{file=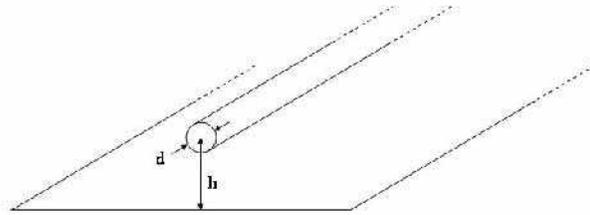}
\caption{Geometry of nanotube in presence of a ground plane.}
\label{fig:geometry}
\end{figure}

\section{Nano-transmission line}

The dc circuit model for a one-channel quantum wire of non-interacting
electrons is well known from the Landauer-Buttiker formalism 
of conduction in quantum systems.  The dc conductance is simply
given by $e^2/h$. If the spin degree of freedom is accounted for,
there are two ``channels'' in a quantum wire: spin up and spin down,
both in parallel. We postpone our discussion of spin until the next
section, and assume for the moment the electrons are spinless.
At ac, the circuit model is not well established experimentally.  
In this manuscript we propose and discuss a transmission line
equivalent circuit model that can be used to predict the dynamical
impedance of a single walled nanotube under a variety of measurement geometries.

The effective circuit diagram we are proposing is shown in
figure~\ref{fig:spinlesscircuit}. Below, we will discuss each of the
four contributions to the total circuit, and then discuss some of its
general properties, such as the wave velocity and characteristic
impedance. For the sake of simplicity, as shown in figure~\ref{fig:geometry},
we will restrict ourselves to the case of a wire over a
``ground plane''.

\subsection{Kinetic Inductance}
In order to calculate the kinetic inductance per unit length, we
follow reference~\cite{BockrathThesis} 
and calculate the kinetic energy per unit length and
equate that with the ${1\over 2}LI^2$ energy of the kinetic inductance. The
kinetic energy per unit length in a 1d wire is the sum of the kinetic
energies of the left-movers and right-movers. If there is a net
current in the wire, then there are more left-movers than
right-movers, say.  If the Fermi level of the left-movers is raised by
$e\Delta\mu/2$, and the Fermi-level of the right-movers is decreased
by the same amount, then the current in the 1d wire is
$I=e^2/h\Delta\mu$. The net increase in energy of the system is the
excess number of electrons ($N = e \Delta \mu /2 \delta$) in the left
vs. right moving states times
the energy added per electron $e \Delta \mu / 2$. Here $\delta$ is the
single particle energy level spacing, which is related to the Fermi
velocity through $\delta = \hbar v_F 2 \pi /L$. Thus the excess kinetic
energy is given by $hI^2/4v_Fe^2$. By equating this energy with the
${1\over 2}LI^2$ energy, we have the following expression for the kinetic
energy per unit length:
\begin{equation}
\label{eq:lkinetic}
{\mathcal L}_K = {h\over 2 e^2 v_F} 
\end{equation}
The Fermi velocity for graphene and also carbon nanotubes is usually
taken as $v_F=8~10^5~m/s$, so that numerically
\begin{equation}
{\mathcal L}_K =  16~nH/\mu m.
\end{equation}

In reference~\cite{BurkeNanotube}, we show that in 1d systems, 
the kinetic inductance will always dominate the magnetic inductance. 
This is an important point for engineering nano-electronics: In
engineering macroscopic circuits, long thin wires are usually
considered to have relatively large (magnetic) inductances. 
This is not the case in nano-wires, where the kinetic inductance dominates.

\subsection{Electrostatic capacitance}
The electrostatic capacitance between a wire and
a ground plane as shown in figure~\ref{fig:geometry} is given by\cite{Ramo}
\begin{equation}
\label{eq:celectrostatic}
{\mathcal C}_E = {2 \pi \epsilon \over
  cosh^{-1}\bigl(2h/d\bigr)}\approx {2\pi\epsilon \over ln(h/d)},
\end{equation}
where the approximation is good to within 1~\% for $h> 2d$.
(If the distance to the ground plane becomes
larger than the tube length such as in some free-standing carbon
nanotubes\cite{Dai}, 
another formula for the capacitance has to be
used, which involves replacing h with the length of the 1d wire.)
This can be approximated numerically as
\begin{equation}
\label{eq:cesnumerical}
{\mathcal C}_E \approx 50~aF/\mu m,
\end{equation}
This is calculated using the standard technique of setting the capacitive energy
equal to the stored electrostatic energy:
\begin{equation}
\label{eq:esenergy}
{Q^2\over 2 C} = {\epsilon\over 2}\int E(x)^2~d^3x,
\end{equation}
and using the relationship between $E$ and $Q$ in the geometry of
interest, in this case a wire on top of a ground plane. 

\subsection{Quantum capacitance}
In a classical electron gas (in a box in 1,2, or 3 dimensions), to add
an extra electron costs no energy. (One can add the electron with any
arbitrary energy to the system.)  In a quantum electron gas (in a box
in 1,2, or 3 dimensions), due to the Pauli exclusion principle it is
not possible to add an electron with energy less than the Fermi energy
$E_F$.  One must add an electron at an available quantum state above $E_F$.
In a 1d system of length L, the spacing between quantum states is
given by:
\begin{equation}
\delta E = {dE\over dk} \delta k = \hbar v_F {2\pi\over L},
\end{equation}
where L is the length of the system, and we have assumed a linear
dispersion curve appropriate for carbon nanotubes.
By equating this energy cost with an effective 
quantum capacitance\cite{BockrathThesis,Tarkiainen}
with energy given by
\begin{equation}
{e^2\over C_Q}=\delta E,
\end{equation}
one arrives at the following expression for the (quantum) capacitance
per unit length:
\begin{equation}
{\mathcal C}_Q = {2 e^2\over h v_F}, 
\end{equation}
which comes out to be numerically
\begin{equation}
{\mathcal C}_Q = 100~aF/\mu m.
\end{equation}

\subsection{Wave velocity}

The wave velocity of a (any) transmission line with
inductance per unit length $\mathcal L$ and capacitance
per unit length $\mathcal C$ is simply $1/\sqrt{{\mathcal L C}}$.
In the case under consideration here, the inductance is simply
the kinetic inductance; the total capacitance per unit length is given by
\begin{equation}
{\mathcal C}_{total}^{-1}={\mathcal C}_Q^{-1} + {\mathcal C}_{ES}^{-1}.
\label{eq:ctotal}
\end{equation}
If we were to neglect the screened coulomb interaction, it would be
equivalent to neglecting the electrostatic capacitance. In that case we have
\begin{equation}
v_{non-interacting} \approx \sqrt{1\over {\mathcal L}_K {\mathcal C}_Q}=v_F.
\end{equation}
One method of including the effect of 
electron-electron interactions in the context of the above discussion
is simply to include the electrostatic capacitance as well as the
quantum capacitance, so that the wave velocity is not quite exactly
equal to the Fermi velocity:
\begin{equation}
v_{interacting} \approx \sqrt{1\over {\mathcal L}_K {\mathcal C}_{total}} = \sqrt{{1\over {\mathcal L}_K {\mathcal C}_{ES}}+{1\over {\mathcal L}_K {\mathcal C}_Q}}>v_F.
\end{equation}
The ratio of the wave (plasmon) velocity in the absence of interactions to
the wave (plasmon) velocity in the presence of interactions has a special
significance in the theory of Luttinger liquids, and is denoted by the
letter ``g''.

\subsection{Characteristic impedance}

Another property of interest of the nano-transmission line is the
characteristic impedance, defined as the ratio of the ac voltage to
the ac current. This is especially important for measurement purposes.
The characteristic impedance of a (any) transmission line with
inductance per unit length $\mathcal L$ and capacitance
per unit length $\mathcal C$ is simply $\sqrt{{\mathcal L/C}}$.

If one considers only the quantum capacitance and
only the kinetic inductance, the characteristic impedance turns out to
be the resistance quantum:
\begin{equation}
Z_{c,non-interacting}=\sqrt{{\mathcal L}_K\over{\mathcal C}_Q}={h\over 2e^2}=12.5~k\Omega.
\end{equation}
If one both considers both components of
the capacitance (electrostatic + quantum), then one finds:
\begin{equation}
Z_{c,interacting}=\sqrt{{\mathcal L}_K\over{\mathcal C}_{total}}
=\sqrt{{{\mathcal L}_K\over{\mathcal C}_{ES}}+{{\mathcal
      L}_K\over{\mathcal C}_Q}}
=g~{h\over 2e^2},
\label{eq:zchar}
\end{equation}
where we have inserted the definition of $g$.

\subsection{Damping Mechanisms?}
\label{sec:damping}
An important question to consider is the damping of the 1d plasma
waves.  Currently very little is known theoretically or experimentally
about the damping {\it mechanisms}. In the absence of such knowledge,
we model the damping as distributed resistance along the length of the
tube, with resistance per unit length $\mathcal R$. 
This model of damping of 2d plasmons we recently
measured\cite{BurkeAPL} was successful in
describing our experimental results, using the dc resistance to
estimate the ac damping coefficient.

\begin{figure}
\epsfig{file=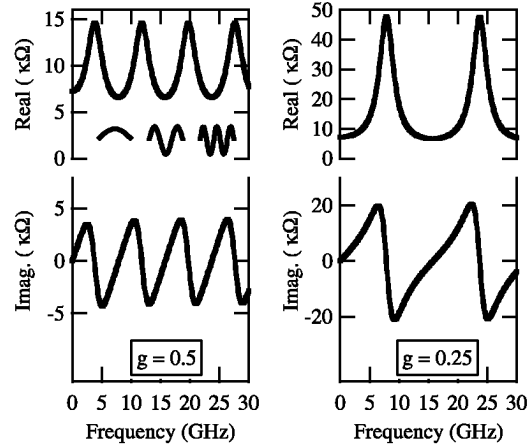}
\caption{Predicted nanotube dynamical impedance for ohmic contact, for two different
  values of g. We assume $l=100~\mu m$, and ${\mathcal
  R}=10~\Omega/\mu m$}
\label{fig:ohmic}
\end{figure}

\section{Spin-charge separation}
\label{sec:spincharge}

A carbon nanotube, because of its band structure, has two propagating
channels\cite{Dresselhaus}. In
addition, the electrons can be spin up or spin down. Hence, there are
four channels in the Landauer-B\"uttiker formalism language. 
In this section we discuss an effective high-frequency
circuit model which includes the contributions of all four channels,
and makes the spin-charge separation (the hallmark of a Luttinger
liquid) clear and intuitive.

The non-interacting ac circuit model (i.e., one neglecting the
electrostatic capacitance) of a single-walled carbon nanotube is fairly
straightforward: One simply has four quantum channels in parallel
each with its own kinetic inductance and quantum capacitance per unit
length. All of the above calculations would apply to that system,
accept that there are four transmission lines in parallel. 

When one includes the effect of electrostatic capacitance, the four
individual propagating modes all share the same capacitance to the
ground plane, and hence become coupled. The equations of motion for
the four coupled transmission lines can be
diagonalized, and the eigenmodes become three spin
waves and one charged (voltage) wave. In
reference~\cite{BurkeNanotube}, we show that the
circuit model of figure~\ref{fig:spinlesscircuit} is still valid as
as an effective circuit model for the charged mode if
the kinetic inductance and quantum capacitance (but {\em not} the
electrostatic capacitance) are simply divided by {\em four}.

\begin{figure}
\epsfig{file=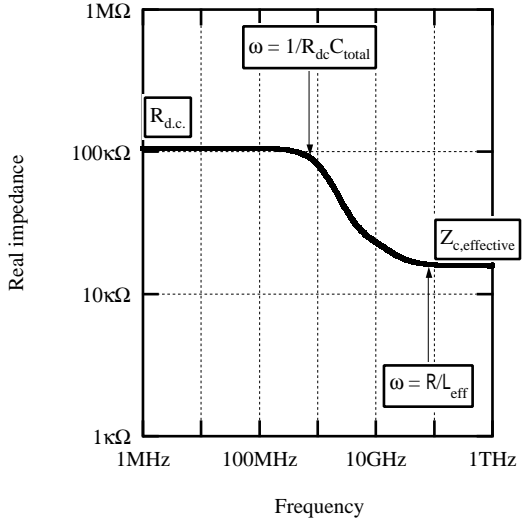}
\caption{Predicted nanotube dynamical impedance in overdamped case.
We assume $l=100~\mu m$, ${\mathcal R}=1~k\Omega/\mu m$, and $g=0.25$.}
\label{fig:overdamped}
\end{figure}

\section{Numerical predictions}

Using the above described circuit description, we numerically
calculate the complex, frequency dependent impedance of an
electrically contacted single
nanotube (including the effects of spin) for two cases: high damping
and low damping, i.e. high and low resistance per unit length.
In the low damping limit, shown in
figure~\ref{fig:ohmic}, the predicted impedance has resonant
frequency behavior, corresponding to standing waves along the length
of the tube. 

In the high-damping limit, shown in figure~\ref{fig:overdamped},
the resonances are washed out and one sees two limits.
First, at dc the real impedance is simply
the resistance per length times the length, i.e. ${\mathcal R}l$, plus
the Landauer-Buttiker contact resistance.  
The frequency scale at which the impedance starts to 
change is given by the inverse of the
total capacitance (${\mathcal C}_{total}l$) times the total
resistance. At very high frequencies, the impedance becomes equal to
the characteristic impedance given in
equation~\ref{eq:zchar} (corrected for the spin-charge effects
discussed in section~\ref{sec:spincharge}). 
The frequency at which this occurs is given by
the inverse of the effective ``LR'' time constant, which is the
resistance per unit length divided by the inductance per unit
length. 

\section{Conclusions}

We have derived an effective rf circuit model for a 
single walled carbon nanotube, including the effects of kinetic
inductance as well as the electrostatic and quantum capacitance.
The nano-transmission line model we developed is a circuit description
of a 1d plasmon, and as such is directly related to the long 
postulated Luttinger liquid properties of 1d systems.
Our next step will be to experimentally test the validity of this circuit
model in either the frequency or time domain, and to develop an rf
circuit model for {\em active} nanotube devices.

\section{Acknowledgments}
This work was supported by the ONR. 
We thank J.P. Eisenstein and Cees Dekker for useful discussions.

%\bibliography{burkenano}

\begin{biography}[\epsfig{file=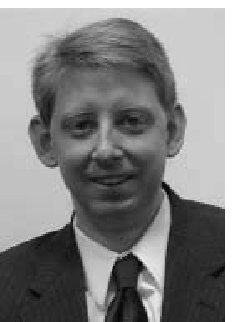}]{Prof. P.J. Burke}
received his Ph.D. in physics from Yale University in 1998.
His thesis research involved a combination of fundamental and applied
physics, aimed at the development of fast Nb superconducting hot-electron
bolometer mixers for sensitive THz heterodyne receivers. From
1998-2001, he was a Sherman Fairchild Postdoctoral Scholar in
Physics at Caltech, where he studied resonant tunnel diodes for THz
detectors and sources, 2d plasmons, and the interaction between Johnson
noise and quantum coherence in 2d systems. Since 2001, he has been an assistant
professor in the department of electrical and computer engineering at
the University of California, Irvine, where, as the
recipient of an ONR Young Investigator Award, he is currently leading
the initiative in nanotechnology.
His research interests include nanoelectronics, nanomechanics, and
nano-biotechnology, 
as well as quantum computation and quantum information processing.
\end{biography}

\end{document}